\documentclass[preprint2]{emulateapj}

\def\nh{N_{\rm H}}
\def\cmmoinsdeux{\textrm{cm}^{-{2}}}
\def\adeg{^{\circ}}
\def\amin{^\prime}
\def\asec{^{\prime \prime}}
\def\cm{\textrm{cm}}
\def\mags{\textrm{mags}}
\def\microns{\mu\textrm{m}}
\def\Al{A_{\rm \lambda}}
\def\Av{A_{\rm V}}
\def\Ak{A_{\rm K}}
\def\igrjstu{\mbox{IGR~J16318$-$4848}}
\def\Tstar{\mbox{ }T_{\star}}
\def\Rstar{\mbox{ }R_{\star}}
\def\Tdust{\mbox{ }T_{\rm dust}}
\def\Rdust{\mbox{ }R_{\rm dust}}
\def\Mdust{\mbox{ }M_{\rm dust}}
\def\Dstar{\mbox{ }D_{\star}}
\def\Rsol{\mbox{ }R_{\odot}}
\def\Msol{\mbox{ }M_{\odot}}
\def\Lsol{\mbox{ }L_{\odot}}
\def\kpc{\mbox{ kpc}}

\slugcomment{to appear in ApJ. June 1, 2012, Issue 751-2}

\shorttitle{Broadband near to mid-infrared spectroscopy of $\igrjstu$}
\shortauthors{Chaty and Rahoui}

\begin{document}


\title{Broadband ESO/VISIR--\textit{Spitzer} infrared spectroscopy of the obscured supergiant X-ray Binary $\igrjstu$}


\author{S. Chaty\altaffilmark{1} and F. Rahoui\altaffilmark{1,2}}
\altaffiltext{1}{AIM (UMR-E 9005 CEA/DSM-CNRS-Universit\'e Paris Diderot)
Irfu/Service d'Astrophysique, Centre de Saclay,
FR-91191 Gif-sur-Yvette Cedex, France} \email{sylvain.chaty@cea.fr}
\altaffiltext{2}{Harvard University, Department of Astronomy \& Harvard-Smithsonian Center for Astrophysics, 60 Garden street, Cambridge, MA 02138, USA} \email{frahoui@cfa.harvard.edu}


\begin{abstract}

A new class of X-ray binaries has been recently discovered by the high energy observatory, {\it INTEGRAL}. It is composed of intrinsically obscured supergiant high mass X-ray binaries, unveiled by means of multi-wavelength X-ray, optical, near- and mid-infrared observations, in particular photometric and spectroscopic observations using ESO facilities. However the fundamental questions about these intriguing sources, namely their formation, evolution, and the nature of their environment, are still unsolved. 
Among them, $\igrjstu$ -- a compact object orbiting around a supergiant B[e] star -- seems to be one of the most extraordinary celestial sources of our Galaxy.
We present here new ESO/VLT VISIR mid-infrared (MIR) spectroscopic observations of this source\footnote{Based on observations made with ESO Telescopes at the La Silla or Paranal Observatories under programme ID \#079.D-0454.}. First, line diagnostics allow us to confirm the presence of absorbing material (dust and cold gas) enshrouding the whole binary system, and to characterise the nature of this material. Second, by fitting broadband near to mid-infrared Spectral Energy Distribution -- including ESO NTT/SofI, VLT/VISIR and \textit{Spitzer} data -- with a phenomenological model for sgB[e] stars, we show that the star is surrounded by an irradiated rim heated to a temperature of $\sim 3800-5500$~K, along with a viscous disk component at an inner temperature of $\sim 750$~K. VISIR data allow us to exclude the spherical geometry for the dust component.
This detailed study will allow us in the future to get better constraints on the formation and evolution of such rare and short-living high mass X-ray binary systems in our Galaxy.
\end{abstract}


\keywords{circumstellar matter --- infrared: stars --- stars: emission-line, B[e] --- X-rays: binaries --- X-rays: individual ($\igrjstu$)}



\section{Introduction} \label{section:introduction}

The high energy {\it INTEGRAL} (INTErnational Gamma-Ray Astrophysics Laboratory) observatory, observing in the 20\,keV$-$8\,MeV range, has performed a detailed survey of the Galactic plane, and the ISGRI detector on the IBIS imager has discovered many new sources, most of them reported in \cite{bird:2010}.
Many of them are concentrated in a direction tangent to the Norma arm of the Galaxy \citep[see e.g.][]{tomsick:2004a, chaty:2008}, rich in star forming regions.  The most important result of the {\it INTEGRAL} observatory to date is the discovery of many new high energy sources exhibiting common characteristics which, previously, had rarely been seen. 
Most of them are High Mass X-ray Binaries (HMXBs) hosting a neutron star orbiting a supergiant O/B companion star. Some of the new sources are very obscured, exhibiting huge intrinsic and local extinction \citep[see e.g.][]{chaty:2011c}. Certainly one of the most extraordinary examples 
among the obscured high energy sources is the extremely absorbed source $\igrjstu$ \citep[see][]{filliatre:2004}.

$\igrjstu$ was the first source to be discovered by ISGRI on 29 January 2003 at the Galactic coordinates $(l,b) \sim (336\adeg, 0.5\adeg$) with an uncertainty radius of localisation of $2\amin$ \citep{courvoisier:2003}. ToO observations were then triggered with {\it XMM-Newton}, allowing a more accurate localisation (RA=16:31:48.6; Dec=-48:49:00) at $4\asec$ \citep{walter:2003}. {\it XMM-Newton} observations showed that the source was exhibiting a strong column density of $\nh \sim 2 \times 10^{24} \cmmoinsdeux$ \citep{matt:2003,walter:2003}. The flux was highly variable (by a factor of 20), with an usual time lapse of 10 hours between flares, and 2 to 3 days of recurrent inactivity. The lines and continuum were varying on a 1000\,s timescale, used to derive the size of the emitting region to be smaller than $3 \times 10^{13} \cm$.  These X-ray properties, signature of wind accretion, were reminiscent of other peculiar high energy sources, such as XTE\,J0421+560/CI\,Cam and GX\,301-2 \citep{revnivtsev:2003c}.

\subsection{Optical and near-infrared observations of $\igrjstu$} \label{section:introductionNIR}

The accurate {\it XMM-Newton} localisation of $\igrjstu$ allowed \cite{filliatre:2004} to trigger ToO photometric and spectroscopic observations in optical and near-infrared (NIR) just after the detection of the source, and to discover the counterpart at NIR wavelengths.  The first striking fact was the extreme brightness of its NIR counterpart, with magnitudes J\,$= 10.33 \pm 0.14$, H\,$=8.33 \pm 0.10$ and K$_{\rm s} = 7.19 \pm 0.05$. The second surprising fact was its absorption.  
The optical/NIR counterpart was exhibiting an unusually strong intrinsic absorption in the optical V-band of $\Av = 17.4 \mags$, much stronger than the absorption along the line of sight exhibited by neighbouring objects ($\Av = 11.4 \mags$), but still 100 times lower than the absorption in X-rays. This led \cite{filliatre:2004} to suggest that the material absorbing in the X-rays had to be concentrated around the compact object, while the material absorbing in optical/NIR would extend around the whole binary system.

The NIR spectroscopy in the $0.95-2.5 \microns$ domain revealed the third amazing fact: the high energy source exhibits an unusual NIR spectrum, very rich in many strong emission lines.  The study of these lines showed that they originated from different media (exhibiting various densities and temperatures), suggesting the presence in this high energy source of a highly complex and stratified circumstellar environment, and also the presence of an envelope and a wind.  Only luminous post main sequence stars show such extreme environments, the companion star being most likely a sgB[e] star, therefore making $\igrjstu$ an HMXB. As in X-rays, the NIR characteristics are also reminiscent of the other peculiar high energy source XTE\,J0421+560/CI\,Cam, for which {\it IRAS} 12-100$\microns$ data suggested the existence of a substantial circumstellar dust shell \citep{belloni:1999,clark:1999}.

By fitting a black body representing the companion star in the multi-wavelength Spectral Energy Distribution (SED) of this source, \cite{filliatre:2004} derived the following parameters: $\Av = 17.5 \mags$, L~$\sim 10^6 D_{\rm 6 kpc}^2 \times L_{\odot}$, T~$= 20250$~K, M~$= 30 M_{\odot}$ and $r/D = 5\times 10^{-10}$, where L, T, M, r and D are the companion star luminosity, photosphere temperature, mass, radius and distance, respectively.  These parameters imply a high luminosity, high temperature, and high mass star, therefore likely a supergiant, located between 1 and 6~kpc. The NIR photometry, spectroscopy and SED fitting all led to the same results.  Finally, by locating these parameters on a Hertzsprung-Russel (or temperature--luminosity) diagram, one can see that this companion star is located at the edge of the blue supergiant domain, indicating that we are facing an extreme object even among blue supergiant stars \citep[see e.g.][]{lamers:1998}.

\subsection{Previous mid-infrared observations of $\igrjstu$}

However, the cause of its unusually strong absorption could not be unveiled by observations in the optical/NIR domain, and only mid-infrared (MIR) data would allow us to characterise the nature of this absorbing material, and determine whether it was made of cold gas, dust, or anything else. \cite{rahoui:2008} obtained MIR photometric observations ($8-19 \microns$) with VISIR on VLT/UT3 of $\igrjstu$. They built the optical to MIR SED by putting together photometric data from ESO NTT/SofI, VISIR and \textit{Spitzer} (GLIMPSE survey). Since data could not be fitted with only the spectral stellar type of a hot luminous sgB[e] star, they added to a model of the companion star (taking usual parameters of a sgB[e]) a simple spherical dust component. They then derived the following parameters: companion star temperature $\Tstar=22000$\,K and radius $\Rstar = 20.4~R_{\odot} = 15 \times 10^6$\,km, and dust component temperature $\Tdust = 1000$\,K and radius $\Rdust = 10 \Rstar = 150 \times 10^6$\,km. The derived absorption and distance were $\Av = 17 \mags$ and D~$=1.6$~kpc respectively, with a fit $\chi^2$/dof of 6.6/6 \citep{rahoui:2008}. The most important result was therefore that they needed an extra component in order to satisfyingly fit the data, whose extension suggested that it was enshrouding the whole binary system, as a cocoon of dust would likely do.

Archival MIR observations taken from the \textit{Spitzer} GLIMPSE survey confirmed the existence of a long-wavelength ($\lambda \geq 4 \microns$) excess in this source \citep{kaplan:2006}. These authors fit the excess with a black body component of temperature $\sim 500-1500$\,K and radius $\sim 10 \Rstar$, consistent with warm circumstellar dust emission \citep[similarly to the XTE J0421+560/CI Cam system,][]{belloni:1999}, and they suggested that this dust could be the cause of the X-ray absorption. However their data could not constrain the properties of the dust (total mass, temperature and density distribution).
Subsequent \textit{Spitzer}-IRS MIR spectroscopic observations by \cite{moon:2007} confirmed the presence of a hot (T$>$700K) circumstellar dust, and also suggested the presence of a warm (T$\sim$190K) dust component, which appeared necessary to get satisfying fits.

Before speculating more on the nature and characteristics of this highly absorbed supergiant X-ray binary, we first need to better characterise the nature and origin of this surrounding dust/cold gas component. This is why we performed MIR spectroscopic observations of this object with VISIR on VLT, currently the most adequate instrument available to perform sensitive MIR spectroscopic observations in the 7-14\,$\microns$ domain, with two primary goals: 

\begin{itemize}

\item to detect emission lines, in order to determine the composition, density and temperature of the absorbing material. 

\item to perform accurate broadband NIR-MIR SED fitting of the stellar and dust component continuum emission of the enshrouded binary system, in order to constrain the temperature, geometry, extension around the system, the nature and composition of the dust component and absorbing material. 

\end{itemize}

\section{Observations and data reduction}

MIR photometric and spectroscopic observations of $\igrjstu$ were carried out on 2007 July 12$-$13 using VISIR \citep{lagage:2004}, the ESO/VLT MIR instrument, composed of an imager and a long slit-spectrometer covering several filters 
in N and Q bands and mounted on Unit 3 (``Melipal'') of the VLT. The standard ``chopping and nodding'' MIR observational technique was used to suppress the
background dominating at these wavelengths. Secondary mirror-chopping was
performed in the North-South direction with an amplitude of $16\asec$ at a frequency of 0.25\,Hz. 
The nodding technique, needed to compensate for chopping residuals, was chosen as
parallel to the chopping and applied using primary mirror offsets of 16\arcsec. 
Because of the high thermal MIR background for ground-based observations, 
the detector integration time was set to 16~ms.

Raw data were reduced using the dedicated IDL reduction package\footnote{\rm http://www.eso.org/sci/facilities/paranal/instruments/visir/doc/VISIR\_datareduction-cookbook\_v080.0.pdf}. The elementary images are co-added in real-time to obtain
chopping-corrected data, then the different nodding positions were
combined to form the final image. The VISIR detector is affected by
stripes randomly triggered by some abnormal high-gain pixels.
A dedicated destriping method was developed to suppress them$^2$.

\subsection{VISIR Photometry} \label{section:VISIRphotometry}

VISIR broadband photometry was performed in six broad and seven narrow filters, whose characteristics are given in Table~\ref{table:photometry}.
Each time we used the small field of view in all bands ($19\farcs2 \times 19\farcs2$ and $0\farcs075$/pixel plate scale). All the observations were bracketed with observations of standard stars \citep{cohen:1999} for flux calibration and PSF determination. We measured the instrumental fluxes and corresponding uncertainties using aperture photometry. They were then converted to physical fluxes using the conversion factors derived from standard stars observations. Results on the photometry are reported in Table~\ref{table:photometry}.
Finally, according to the ESO manual\footnote{www.eso.org/sci/facilities/paranal/instruments/visir/doc}, systematic uncertainties of VISIR photometric data are of the order of 10\% in PAH1 and PAH2, and 20\% in Q2.


\subsection{VISIR Spectroscopy} \label{section:VISIRspectroscopy}

\begin{figure}[!ht]
	\begin{center}
	\includegraphics[angle=90,width=17.cm]{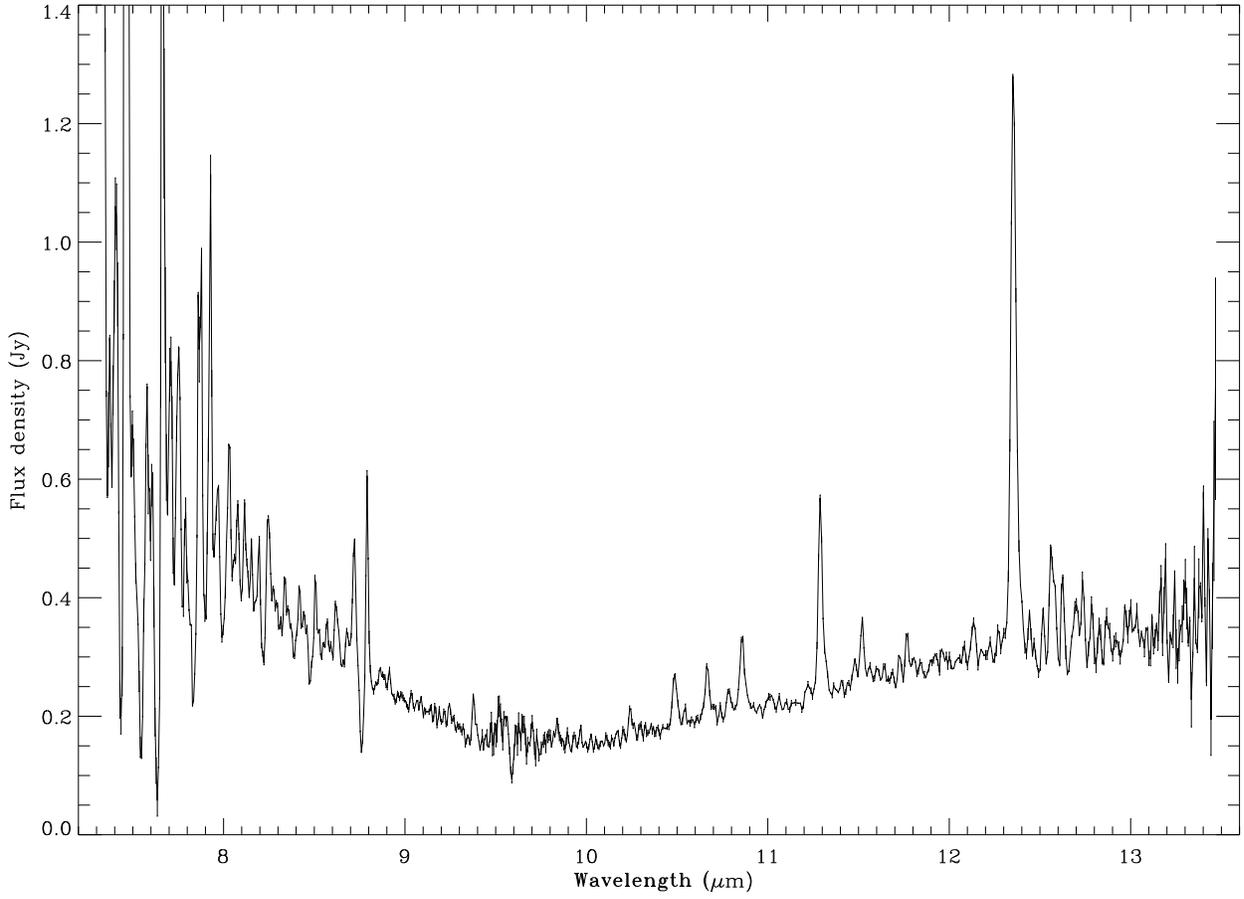}
	\end{center}
	\caption{
Broadband ESO/VLT VISIR MIR spectrum of $\igrjstu$ from 7.2 to $13.6~\microns$. Particularly visible is the broad and prominent silicate absorption feature around $9.7\microns$, due to dust component in the interstellar medium.
%
}
	\label{figure-spectrumMIR}
	\end{figure}

\begin{figure}[!ht]
	\begin{center}
	\includegraphics[angle=90,width=17cm]{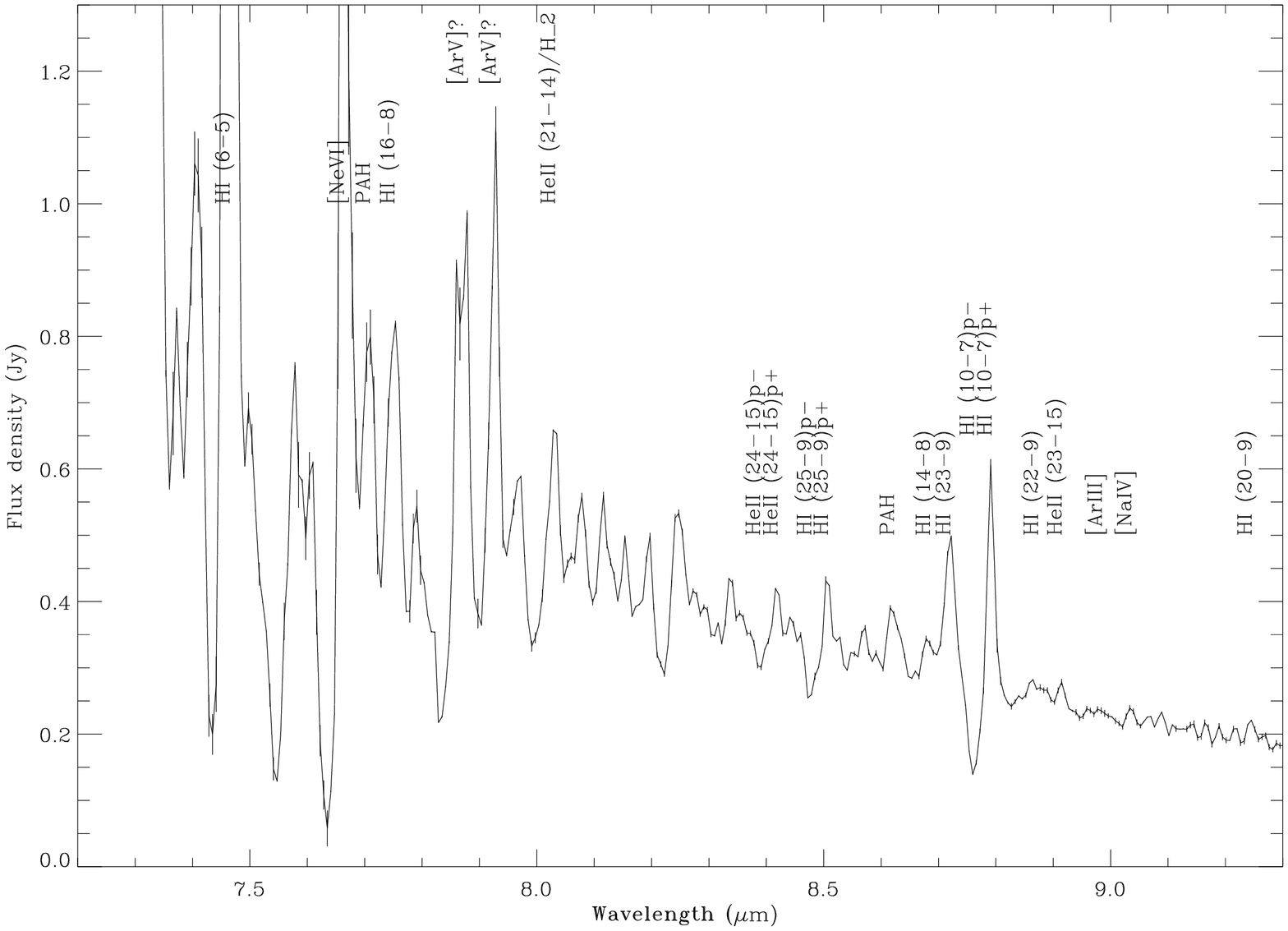}
	\end{center}
	\caption{
Enlarged ESO/VLT VISIR MIR spectrum of $\igrjstu$ from 7.2 to $9.3~\microns$ with identified lines indicated.
}
	\label{figure-spectrumMIR_7-9}
	\end{figure}

\begin{figure}[!ht]
	\begin{center}
	\includegraphics[angle=90,width=17cm]{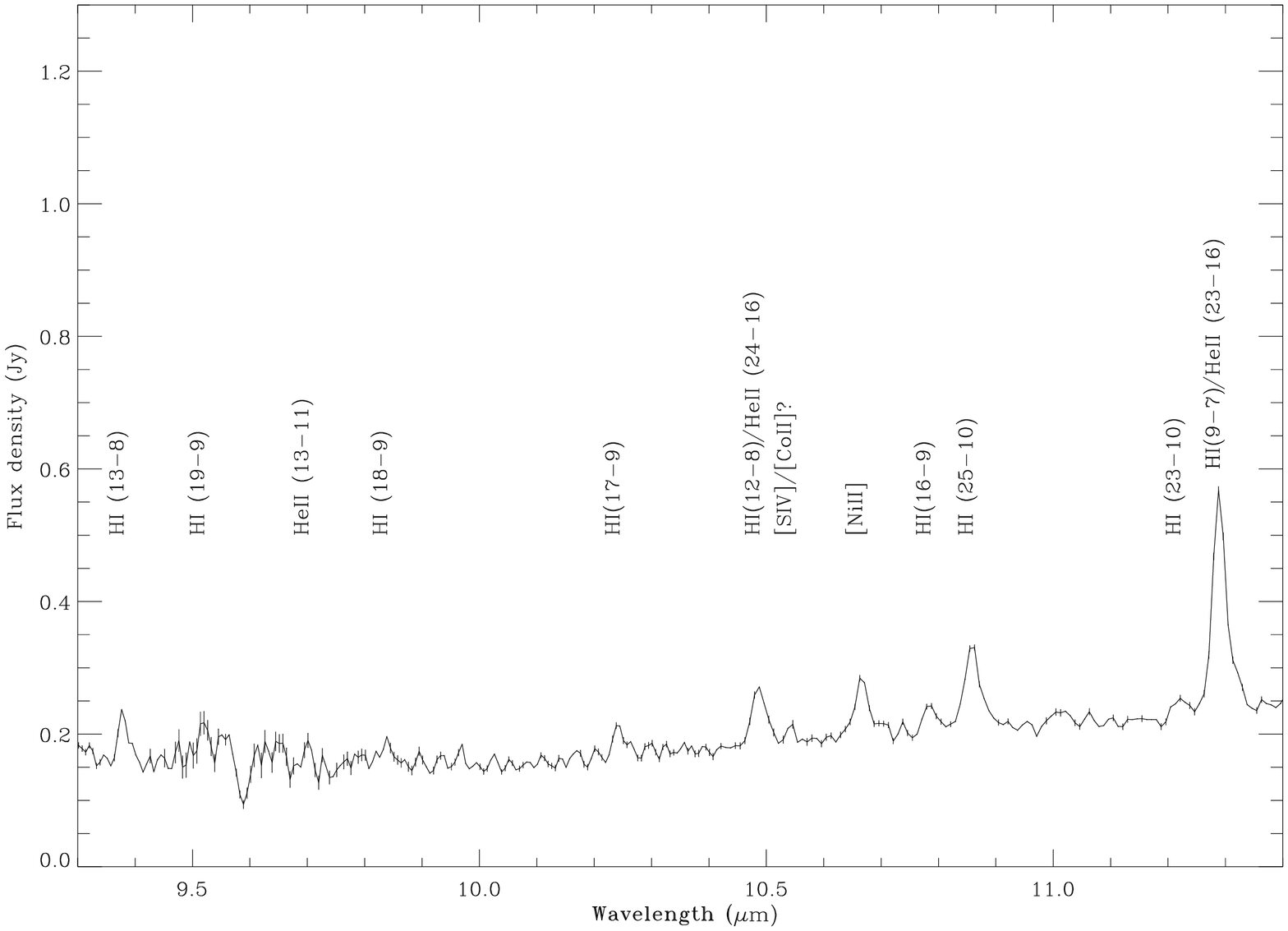}
	\end{center}
	\caption{
Enlarged ESO/VLT VISIR MIR spectrum of $\igrjstu$ from 9.3 to $11.4~\microns$ with identified lines indicated.
}
	\label{figure-spectrumMIR_9-11}
	\end{figure}

\begin{figure}[!ht]
	\begin{center}
	\includegraphics[angle=90,width=17cm]{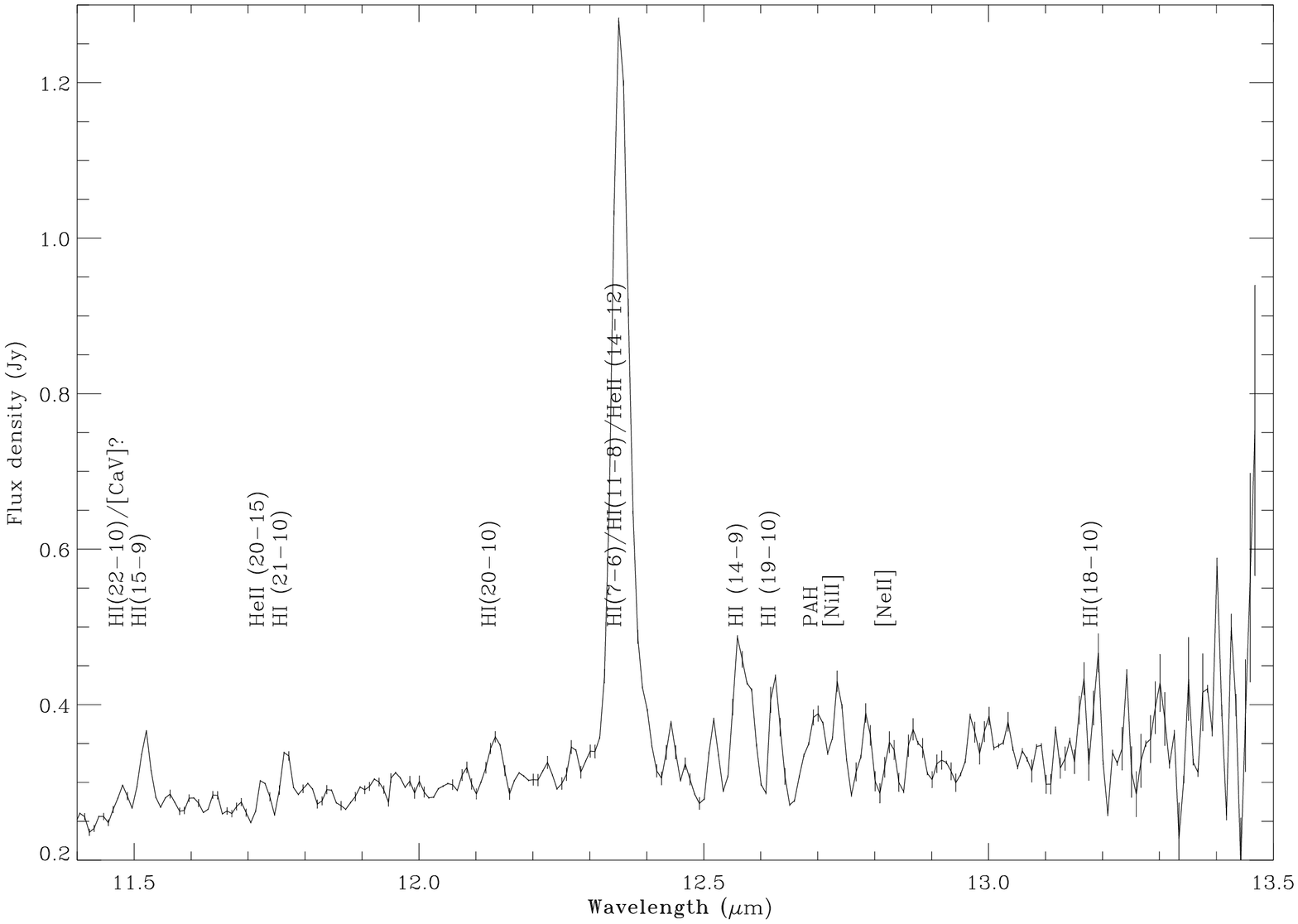}
	\end{center}
	\caption{
Enlarged ESO/VLT VISIR MIR spectrum of $\igrjstu$ from 11.4 to $13.5~\microns$ with identified lines indicated.
}
	\label{figure-spectrumMIR_11-13}
	\end{figure}

We also performed VISIR low-resolution spectroscopy ($R = \frac{\lambda}{\Delta \lambda} \approx350$) of $\igrjstu$, from $7.1$ to $13.46 \microns$ split in five filters of the N-band, centered at $8.1$, $8.8$, $9.8$, $11.4$ and $12.4 \microns$. Exposure times were equal to 1\,hour for the first filter and 2\,hours for all the others, and we reached S/N from about 17 to 56.
During the same night and in the same conditions, we also observed the standard star HD~149447 \citep{cohen:1999}, a K6\,III giant. Extraction of the spectra, wavelength calibration, telluric lines correction, and flux calibration were then carried out using the IDL reduction package mentioned above. In addition, in order to accurately flux calibrate the spectra, they have been scaled to match the VISIR PAH2 photometric flux.
The VISIR MIR broadband spectrum is shown in Figure \ref{figure-spectrumMIR}, zooms on different parts of the spectrum are shown in Figures \ref{figure-spectrumMIR_7-9}, \ref{figure-spectrumMIR_9-11} and \ref{figure-spectrumMIR_11-13}, respectively. The detected lines are indicated in the Figures, and reported in Table~\ref{table:spectroscopy}. 


\subsection{Archival data}

With the aim of SED fitting, we completed our set of VISIR data with already published (1) ESO NTT/SofI NIR \citep{filliatre:2004} and (2) \textit{Spitzer}/IRS MIR \citep{moon:2007} low-resolution spectra. We have reanalysed and recalibrated both spectra:

(1) The $\igrjstu$ and HiP~80456 (F5V) telluric standard SofI spectra, covering the spectral range $0.9$ to $2.5 \microns$, were reduced with the \textit{IRAF} suite by performing crosstalk correction, flatfielding, sky subtraction, and bad pixel correction. The spectra were then extracted, wavelength calibrated with a xenon arc taken with the same setup, and finally combined. 
The telluric features were corrected with the \textit{telluric} task. The result was then multiplied by an F5V synthetic spectrum downloaded from the ESO website\footnote{http://www.eso.org/sci/facilities/paranal/instruments/isaac/tools/lib/index.html}, scaled to the 2MASS magnitudes of HiP~80456 in the H and K filters. The final $\igrjstu$ SofI spectrum is therefore flux calibrated with about 5\% uncertainties on the continuum level.

(2) We used the SL2 ($5.20-7.70 \microns$), SL1 ($7.40-14.50 \microns$), LL2 ($14.00-21.30 \microns$), and LL1 ($19.50-38.00 \microns$) low-resolution \textit{Spitzer} spectra of $\igrjstu$. Basic Calibration Data (BCD) were reduced following the standard procedure given in the IRS Data Handbook\footnote{http://ssc.spitzer.caltech.edu/irs/irsinstrumenthandbook/IRS\_Instrument\_Handbook.pdf}. 
The basic steps were bad pixel correction with {\tt IRSCLEAN} v1.9, sky subtraction, as well as 
extraction and calibration (wavelength and flux) of the spectra $-$ with the \textit{Spectroscopic Modeling Analysis and Reduction Tool} software {\tt SMART} v8.1.2 $-$ for each nodding position. Spectra were then nod-averaged to improve the S/N ratio.

The broadband NIR to MIR ESO NTT/SofI, VISIR and \textit{Spitzer} spectra are shown in Figure \ref{figure:spectrumALL}.

\subsection{Absorption on the line-of-sight and dereddening}

The VISIR spectrum of $\igrjstu$ shows a strong absorption feature, around $9.7 \microns$, most likely due to silicate dust in the interstellar medium (ISM) along the line-of-sight. As described in Section \ref{section:introduction}, several consistent measurements of the visible extinction $\Av$ are found in the literature, most of them derived from SED fitting, and all give values between 17 and $19 \mags$ \citep{filliatre:2004, moon:2007, rahoui:2008}. The silicate absorption feature due to diffuse ISM is strongly correlated to the optical extinction following the relation $\Av\,=\,(18.5\pm2)\times \tau_{\rm 9.7}$ \citep{draine:2003}, where $\tau_{\rm 9.7}$ is the optical depth of the silicate absorption at $9.70 \microns$. It is therefore possible to derive it from the VISIR spectrum. Following \citet{chiar:2007}, we took $\tau_{\rm 9.7}\,=\,-ln(F_{\rm 9.7}/F_{\rm continuum})$, where $F_{\rm 9.7}$ and $F_{\rm continuum}$ are the source and continuum fluxes at $9.70 \microns$ respectively.  The continuum was fitted using both ranges $5.20-7.00 \microns$ and $13.00-14.50 \microns$ (in order to exclude the contribution of the silicate absorption feature) with a second order polynomial. We point out that this method is strongly uncertain, especially concerning the continuum fitting, and the result should therefore be considered with caution. We nevertheless derive $\Av\,=\,18.3\pm0.4 \mags$, in agreement with the previous determinations.

We used this value to deredden the spectra with the extinction laws given in \citet{chiar:2006}. In their paper, these authors derived the $\Al/\Ak$ ratio rather than the usual $\Al/\Av$ one. To express the extinction ratio in a standard way, we assigned to $\Ak$ the value derived from the law for the diffuse ISM \citep[$R_{\rm V}\,=\,3.1$,][]{fitzpatrick:1999a}: $\Ak\,=\,0.111\times \Av$. Moreover, these authors find that beyond $8 \microns$, the MIR extinction and silicate absorption features are best described by two methods, using: 1) the diffuse ISM (which stops at $27 \microns$), and 2) the Galactic center. We used both cases to deredden our spectra of $\igrjstu$, however the Galactic center case clearly overestimates the silicate absorption depth. We therefore adopt the corrections based on the extinction due to the diffuse ISM.


\section{Results}

\subsection{Photometry}

We first compare in Table \ref{table:photometry} $\igrjstu$ MIR photometric fluxes with our two previous VISIR observations \citep{rahoui:2008}, to see whether there is variability in the MIR emission of this source.
By examining these values, it appears that there exists a small variation of the fluxes:

\begin{itemize}

\item the PAH1 flux has increased by $\sim 6-7 \sigma$ between 2005 and 2006, then remained constant taking into account the error bars between 2006 and 2007.

\item the PAH2 flux has decreased by $1.5 \sigma$ between 2005 and 2006, then further decreased by $\sim 8 \sigma$ between 2006 and 2007.

\item the Q2 flux remained constant taking into account the error bars between 2005 and 2006, then increased by $\sim 2 \sigma$ between 2006 and 2007.

\end{itemize}

There are therefore 2 significant ($> 6 \sigma$) variations of the flux: first the increase of the PAH1 flux between 2005 and 2006, then the decrease of the PAH2 flux between 2006 and 2007. All the other variations are not significant ($< 2 \sigma$). Considering the systematic uncertainties given in Section \ref{section:VISIRphotometry}, we consider in this paper that these variations are within the error range and that the source remains at a stable flux. We accordingly use this photometry to better calibrate the spectra, as described in Section \ref{section:VISIRspectroscopy} and shown in Figure \ref{figure:spectrumALL}, in order to accurately compute fits of $\igrjstu$ SED. Future additional photometric observations should allow us to better constrain the potential flux variability on a long timescale. 

\subsection{Spectroscopy} \label{section:spectroscopy}

With the spectroscopic observations, we first confirm the extremely rich and complicated MIR environment of the source already reported by \cite{moon:2007}, however we detect many more emission lines in our MIR spectra obtained with the VISIR instrument. 
In these spectra, presented in Figures \ref{figure-spectrumMIR}, \ref{figure-spectrumMIR_7-9}, \ref{figure-spectrumMIR_9-11}, \ref{figure-spectrumMIR_11-13} and also \ref{figure:spectrumALL}, we clearly see the following lines, reported in detail in Table \ref{table:spectroscopy}:

\begin{itemize}

\item the broad and prominent silicate absorption feature around $9.7\microns$, due to dust component in the interstellar medium \citep{draine:2003};

\item numerous \ion{H}{1} recombination lines -- the most prominent being \ion{H}{1}(7-6) at $12.35\microns$ -- due to ionized stellar winds;

\item many forbidden atomic and ionic fine-structure lines of low ionization potential in emission -- [\ion{Ne}{2}], [\ion{Ne}{6}], [\ion{Na}{4}], [\ion{S}{4}], [\ion{Ar}{3}], [\ion{Ar}{5}], [\ion{Ca}{5}], [\ion{Ni}{2}], and tentatively [\ion{Co}{2}]--, caused by extended ionized regions of low density;


\item many higher ionization \ion{He}{2} emission lines;

\item and various polycyclic aromatic hydrocarbon (PAH) emission lines, originating from photodissociated regions.

\end{itemize}

Some lines, especially H and He emission lines, exhibit P Cygni profile. The presence of many forbidden lines of low excitation metals such as [\ion{Ni}{2}] indicates that the circumstellar material is geometrically extended, with a large amount of low density gas \citep{lamers:1998}. Most of the lines that we report exhibit the same fluxes as in \cite{moon:2007}, apart from \ion{H}{1}(16-9) and [\ion{Ni}{2}] (fainter by a factor 2) and the [\ion{Ne}{2}] $12.8 \microns$ line (ten times fainter). In the spectrum of \cite{moon:2007} there is a line at the position of [\ion{Ne}{2}] which is much stronger than that in our spectrum, however since their spectral resolution is lower than ours, it is difficult to compare fluxes of these lines. On the other hand, X-ray emission coming from the compact object might photoionize and increase the [\ion{Ne}{2}] $12.8 \microns$ line flux, and not higher excitation states. 

We can use various lines of the MIR domain as indicators of the physical medium state of the source, sensitive to the temperature of the environment, based on the work of \cite{furness:2010} (to our knowledge there is no other line diagnostics that we could use in this MIR domain). Since for most of the lines there is no significant variation, we also use for these diagnostics two additional lines coming from \cite{moon:2007}: [\ion{Ne}{3}] $15.56 \microns$ and [\ion{S}{3}] $18.71 \microns$. We therefore find that:

\begin{itemize}

\item The intensity ratio, $log(\frac{[{Ar\,III}]~8.99 \microns}{[{Ne\,II}]~12.81 \microns}) = log (\frac{2.02e-21}{2.16e-21}) = -0.029$, indicates an effective temperature T$_{\rm eff} \sim 38000-40000$\,K.

\item The intensity ratio, $log(\frac{[{Ne\,III}]~15.56 \microns}{[{Ne\,II}]~12.81 \microns}) = log (\frac{2.60e-21}{2.16e-21}) = 0.08$, indicates an effective temperature T$_{\rm eff} \sim 38000-40000$\,K. This intensity ratio correlates well with $log(\frac{[{Ar\,III}]~8.99 \microns}{[{Ne\,II}]~12.81 \microns})$, taking into account the scattering. If instead we take the value of [\ion{Ne}{2}]~$12.81 \microns$ given in \cite{moon:2007}, and therefore obtained at the same time than [\ion{Ne}{3}]~$15.56 \microns$, we derive a value of $\sim 35000-36000$\,K.

\item The intensity ratio, $log(\frac{[{S\,IV}]~10.52 \microns}{[{Ne\,II}]~12.81 \microns}) = log (\frac{1.11e-21}{2.16e-21}) = -0.29$, indicates an effective temperature T$_{\rm eff} \sim 37000-38000$\,K. This ratio agrees well with $log(\frac{[{Ne\,III}]~15.56 \microns}{[{Ne\,II}]~12.81 \microns})$, taking into account the scattering.

\item The intensity ratio, $log(\frac{[{S\,IV}]~10.52 \microns}{[{S\,III}]~18.71 \microns}) = log (\frac{1.11e-21}{7.73e-21}) = -0.84$, is close to the correlation with $log(\frac{[{Ne\,III}]~15.56 \microns}{[{Ne\,II}]~12.81 \microns})$, taking into account the scattering and uncertainties.

\item the $log (\eta <S-Ne>)$ versus $log(\frac{[{Ne\,III}]~15.56 \microns}{[{Ne\,II}]~12.81 \microns})$ corresponds well to an effective temperature T$_{\rm eff} \sim 38000$\,K with $log U = -1$, adopting the same notations than in \cite{furness:2010}, in particular $\eta <S-Ne> = (\frac{[{S\,IV}]~10.52 \microns}{[{S\,III}]~18.71 \microns}) / (\frac{[{Ne\,III}]~15.56 \microns}{[{Ne\,II}]~12.81 \microns})$.

\end{itemize}

As discussed later in Section \ref{section:discussion}, this $\sim 37000-40000$\,K temperature --higher than a $\sim 20000$\,K photosphere of a BI star-- corresponds to a heated zone surrounding the stellar photosphere.

\subsection{Spectral energy distribution fitting} \label{section:SED}

The presence of PAH and fine-structure emission lines such as [\ion{Ne}{2}] or [\ion{Ar}{3}] proves the existence of a photoionized dust component intrinsic to $\igrjstu$. Moreover, \citet{kaplan:2006} and \citet{rahoui:2008} detected strong NIR and MIR excesses, usual in sgB[e] stars \citep{lamers:1998}, that they fitted with a $\approx1000$~K spherical black body, likely corresponding to warm circumstellar dust, with an extension of $\sim 10 \Rstar$. Nevertheless, this modeling is too simplistic, for the two following reasons. First because, as shown by \citet{moon:2007}, by fitting the spectral continuum from 5 to $35 \microns$ instead of photometric fluxes, a cold $\sim 190$~K dust component is needed in addition to the warm one. Second because, if the $\igrjstu$ companion star is indeed a sgB[e] as shown by \cite{filliatre:2004}, the dust responsible for the NIR/MIR excess should thereby be distributed in an equatorial disk-like structure, instead of presenting a spherical geometry \citep[see e.g.][]{lamers:1998}. 

\begin{figure}[!ht]
	\begin{center}
	\includegraphics[angle=0,width=17cm]{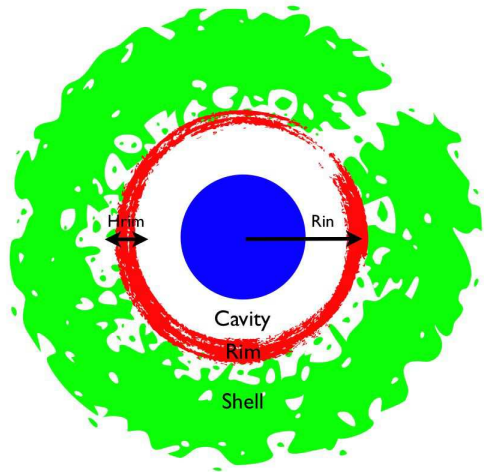}
	\end{center}
	\caption{
Representation of the sgB[e] scheme, with the 3 components seen from polar view (this diagram is just for illustration purpose, and is not to scale, see details in Section \ref{section:SED}).
}
	\label{figure:sgBe}
	\end{figure}

Recently, several improvements have been achieved in the understanding of the emission of intermediate-mass stars such as Herbig Ae/Be stars (HAEBE). Besides the star, these systems are known to exhibit an equatorial disk composed of (see Figure~\ref{figure:sgBe}):
{\it i.} an inner dust-free cavity filled by optically thin gas, 
{\it ii.} a hot and optically thick irradiated rim made of puffed-up hot dust at the sublimation temperature -- responsible for the NIR excess--, and
{\it iii.} a surrounding component of warm dust shell responsible for the MIR excess \citep[see e.g.][and references therein]{dullemond:2001, monnier:2005}.

This scheme was simplified by \citet{lachaume:2007}, by assuming that the disk was completely flat, with a ring-like rim of constant temperature $T_{\rm rim}$, located at an inner radius $R_{\rm in}$ and of width $H_{\rm rim}$ (see Figure~\ref{figure:sgBe}). The remaining parts of this disc are located at an inner radius $R_{\rm in}+H_{\rm rim}/2$, modelled with a radial temperature profile $T\left(R\right)$ as given below in Equation~\ref{sedeq}.
In the case of HAEBE, the rim is assumed to be located at the dust sublimation radius, with a temperature of 1500$-$2000~K, in order to put a constraint on $R_{\rm in}$. In the case of sgB[e] stars the presence of strong \ion{O}{1} emission lines detected in their NIR spectra, as it is the case for  $\igrjstu$ \citep{filliatre:2004}, rather points towards a temperature between $\sim 5000$ and $10000$~K \citep{kraus:2007}. Therefore, since there is no constraint on the disk inner radius $R_{\rm in}$ value, it will be a free parameter, to make the fit more accurate. We adopt the emission model summarised in Equation~\ref{sedeq} to fit $\igrjstu$ SED:

\begin{eqnarray}
F_\nu\, = \,\left ( \frac{R_\ast}{D_\ast} \right )^2\times B(\nu,T_\ast)+2\frac{H_{\rm rim}R_{\rm in}}{{D_\ast}^2}cos\left(i\right)B(\nu,T_{\rm rim}) \\
+ 2\frac{cos\left(i\right)}{{D_\ast}^2}\int_{R_{\rm in}}^{R_{\rm out}} RB(\nu,T\left(R\right))dR\nonumber \\
\rm{with}~T(R)\, = \,T_{\rm in}(R/R_{\rm in})^{-q}
\label{sedeq}
\end{eqnarray}

where $B$ is the black body function at the frequency $\nu$, $T_\ast$, $R_\ast$ and $D_\ast$ the star temperature, radius and distance respectively, and $i$, $T_{\rm in}$, $R_{\rm out}$ and $q$ the disk inclination angle, inner region temperature, outer radius and a dimensionless ``temperature index'' parameter respectively. $T$ and $R$ are the generic temperature and radius respectively.

We can re-write Eq.~\ref{sedeq} using the new variables $r_\ast=\frac{R_\ast}{D_\ast}$, a generic radius $r=R\frac{\sqrt{cos(i)}}{D_\ast}$, $r_{\rm in}=R_{\rm in}\frac{\sqrt{cos(i)}}{D_\ast}$, $r_{\rm out}=R_{\rm out}\frac{\sqrt{cos(i)}}{D_\ast}$, $h_{\rm rim}=H_{\rm rim}\frac{\sqrt{cos(i)}}{D_\ast}$, and we obtain:
\begin{eqnarray}
F_\nu\,&=&\,{r_\ast}^2 B(\nu,T_\ast)+2h_{\rm rim}r_{\rm in} B(\nu,T_{\rm rim}) \\ 
+ 2\int_{r_{\rm in}}^{r_{\rm out}} rB(\nu,T\left(r\right))dr\nonumber\\
\rm{with}~T(r)\,&=&\,T_{\rm in}(r/r_{\rm in})^{-q}
\label{sedeq2}
\end{eqnarray}

During the fitting process, we fixed $T_\ast=20000$~K (consistent with the temperature of a sgB[e] as derived by \citeauthor{filliatre:2004} \citeyear{filliatre:2004}, \citeauthor{rahoui:2008} \citeyear{rahoui:2008}) and considered two cases for the temperature index parameter: $q=0.5$ for an irradiated/flaring disk \citep{chiang:1997}, and $q=0.75$ for a viscous disk \citep{shakura:1973}.

We first fit the data using the flaring disk model, however in each case but one (reported in Table~\ref{table:fit}), $T_{\rm rim}$ is found to be lower than the inner dusty disk temperature $T_{\rm in}$, and we can not obtain any constraint on a parameter set corresponding to high $T_{\rm rim}$. This result does not appear physical since it is not consistent with sgB[e] characteristics, and it likely means that the flaring disk model is not adapted for a sgB[e] star.

We then fit the data using the viscous disk model, and the best fit is obtained for an irradiated rim of temperature $T_{\rm rim} = 3786$\,K, along with a hot dusty viscous disk component at $T_{\rm in} = 767$~K. The derived best-fit parameters are listed in Table~\ref{table:fit}, and the broadband NIR to MIR resulting fit on ESO NTT/SofI, VISIR and \textit{Spitzer} data is reported in Figure \ref{figure:spectrum-fit-best}.

\begin{figure}[!ht]
	\begin{center}
	\includegraphics[angle=0,width=17cm]{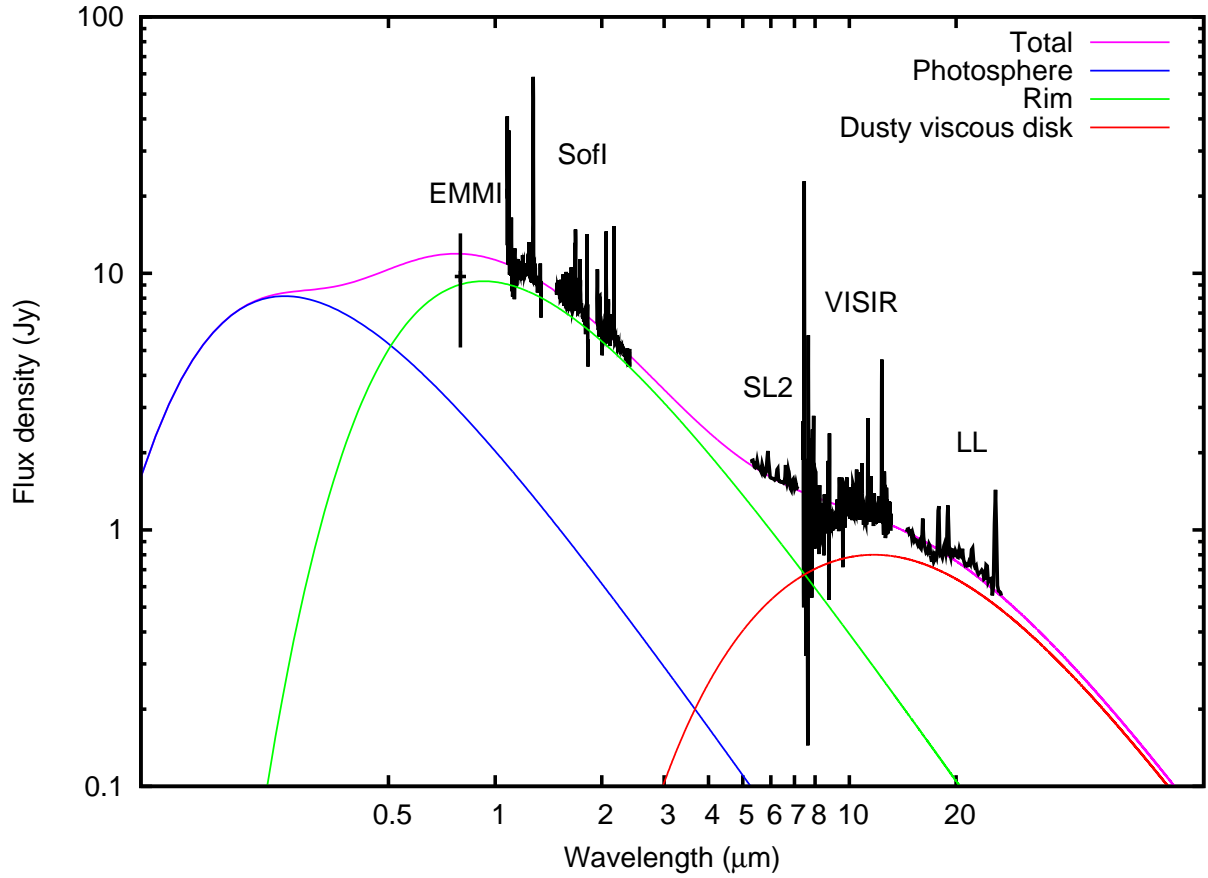}
	\end{center}
	\caption{
Broadband NIR to MIR fit on ESO NTT/SofI, VISIR and \textit{Spitzer} {\it dereddened} spectrum of $\igrjstu$.
SED with a stellar spectral type corresponding to a sgB[e] with $T_\ast=20000$~K, $T_{\rm rim} = 5500$\,K and dust temperature of $\sim 895$\,K. This is not the best fit, however it seems physically more real to describe a sgB[e] star (see text, section \ref{section:SED}).
}
	\label{figure:spectrum-fit-visc-5500}
	\end{figure}

However, since this value seems quite low for a sgB[e] star exhibiting \ion{O}{1} emission lines (see above), we tried various additional fits by freezing reasonable $T_{\rm rim}$ going from 4000 to 6000\,K. For $T_{\rm rim}$ higher than 5963\,K, the stellar contribution is not necessary anymore, which is very likely unreal, since we clearly see its contribution in the NIR domain. We also point out that all the fits with $T_{\rm rim}$ between 3786\,K and 5500\,K pass through the I band data point (taking into account the error bar), which is not the case for $T_{\rm rim}$ higher than this. This is another argument in favour of $T_{\rm rim}$ in this range. For these cases, the most satisfying fit seems to be reached for an irradiated rim of temperature $T_{\rm rim} \sim 5500$\,K, along with a hot dusty viscous disk component at $T_{\rm in} = 895$~K. The derived parameters are listed in Table~\ref{table:fit}, and the broadband NIR to MIR resulting fit on ESO NTT/SofI, VISIR and \textit{Spitzer} data is represented in Figure \ref{figure:spectrum-fit-visc-5500}.

Finally, we also compare fits using i. SofI-VISIR-\textit{Spitzer} data and ii. SofI-\textit{Spitzer} data (i.e. without VISIR data).
First, by taking components of spherical geometry only, both fits exhibit similar goodness, with SofI-VISIR-\textit{Spitzer} data fits leading to a slightly smaller reduced $\chi^2$ of 1.29, compared to 1.32 for SofI-\textit{Spitzer} data, and best fit dust temperatures of 2668-3108 and 379-412\,K for both spherical components respectively (see Tables \ref{table:fit} and \ref{table:fit-Spitzer} at the ``2-sphere component'' line). In addition to a high value of the reduced $\chi^2$, it is the temperature of the first spherical component, too high to be consistent with any kind of dust, which tends to exclude the spherical alternative.
Second, and this is the most interesting result, by taking components of disk-like geometry, the fit is significantly better with SofI-VISIR-\textit{Spitzer} data (reduced $\chi^2$ of 1.19 compared to 1.30 for SofI-\textit{Spitzer} data, with best fit temperatures of 3786 and 767\,K for both components respectively), probably due to the fact that more data points are available in this crucial wavelength range. Therefore, while \textit{Spitzer}-only data do not allow us to distinguish between the two geometries, adding VISIR data lead to a quantitative fit clearly in favour of the disk-like geometry.
By excluding the spherical geometry for the irradiated component, VISIR data are thus essential for a better assessment of the dust component geometry surrounding $\igrjstu$.

\section{Discussion and perspectives} \label{section:discussion}

The multi-wavelength observations that we have performed from optical to MIR on the highly absorbed HMXB $\igrjstu$ hosting a supergiant B[e] star allow us to strengthen the presence of absorbing material (dust and cold gas) enshrouding the whole binary system and better characterise its properties. With these new MIR spectroscopic observations and NIR-MIR ESO NTT/SofI, VISIR and \textit{Spitzer} spectral fitting we confirm the presence of two-temperature components, with an irradiated rim of temperature $T_{\rm rim} = 3786$\,K, surrounded by a hot dusty viscous disk component at $T_{\rm in} = 767$~K, corresponding to the best fit parameters (see Table~\ref{table:fit}).

As suggested by \cite{filliatre:2004} and then shown by \cite{kaplan:2006}, \cite{moon:2007} and \cite{rahoui:2008}, our MIR spectra and SED fitting confirm the existence of a strong NIR/MIR excess, due to the presence of a hot circumstellar dust component, likely formed in the dense outflows coming from the early-type sgB[e] companion star of $\igrjstu$. This component has a temperature of $\Tdust \sim 767-923$~K, with a maximum extension of $r_{\rm out} = 3.55$~au/kpc (assuming $cos(i) = 1$) for $\Rstar / \Dstar$ between $5.80-11.71 \Rsol$\,/kpc. Assuming a distance of $1.6$\,kpc \citep{rahoui:2008}, $\Rstar$ would be in the range $9.3-18.7 \Rsol$, and the extension of the dust equal to $r_{\rm out} = 5.6$\,au, corresponding to $64.3-129.8 \Rstar$. This value is larger than the one derived with the spherical shell model, which is consistent with the fact that for the same amount of dust, it will have to be spread on a larger distance if it has a disk rather than shell geometry. In addition, a high value is consistent with a low inclination, as suggested by the Br$\gamma$ line profile in the NIR spectrum \citep{filliatre:2004}.

On the other hand, in none of our NIR-MIR fits do we require a warm $\sim 190$~K dust component that was claimed by \cite{moon:2007}. We suggest that this additional component was necessary in their fits, firstly because they only took into account the \textit{Spitzer} spectrum without including the NIR spectrum, and secondly because they chose a spherical geometry. By including all available spectra from NIR to MIR, and applying a disk-geometry to our components, our fits are much closer to the physical nature of the environment surrounding these objects, and therefore more accurate and physical.

\begin{figure}[!ht]
	\begin{center}
	\includegraphics[angle=0,width=17cm]{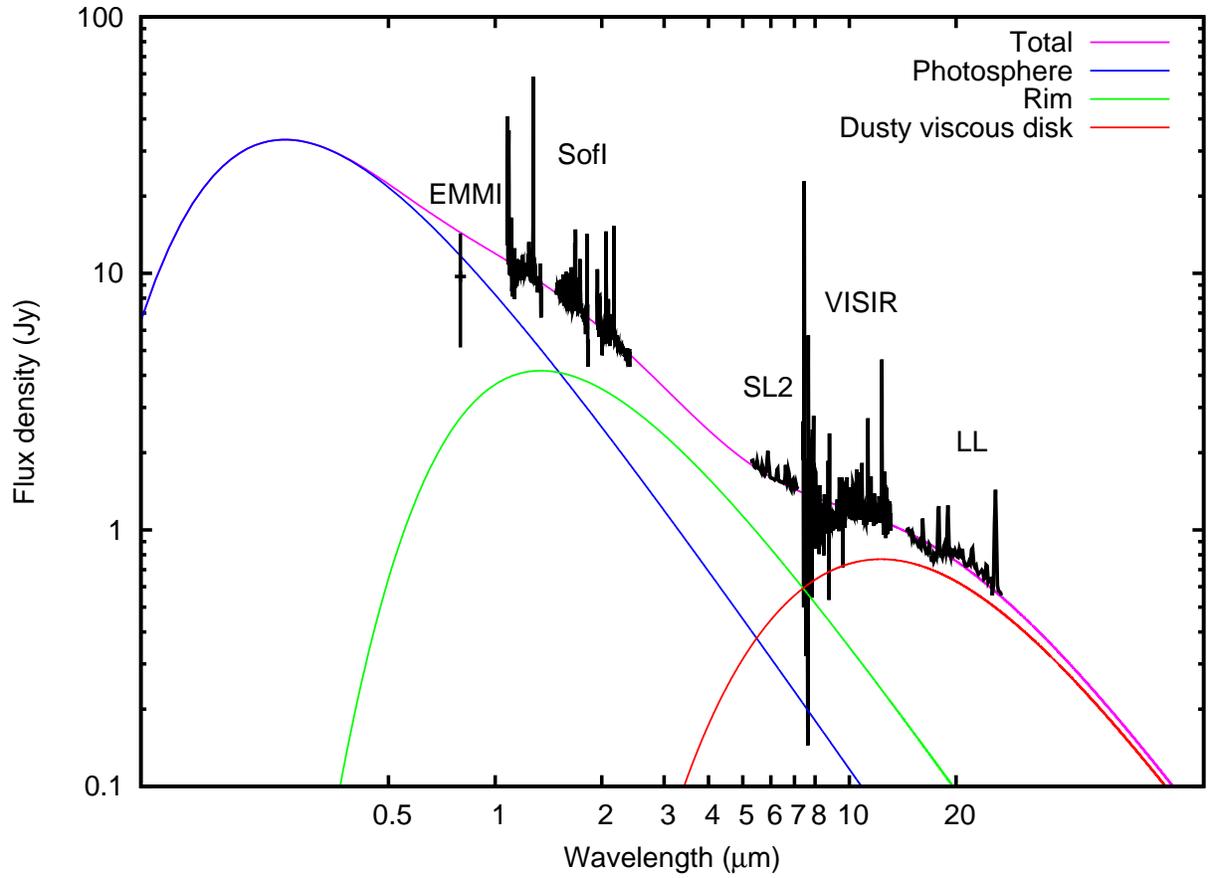}
	\end{center}
	\caption{
Broadband NIR to MIR fit on ESO NTT/SofI, VISIR and \textit{Spitzer} {\it dereddened} spectrum of $\igrjstu$.
SED with a stellar spectral type corresponding to a sgB[e] with $T_\ast=20000$~K, $T_{\rm rim} = 3786$\,K and dust temperature of $\sim 767$\,K. These parameters correspond to the best fit, however $T_{\rm rim}$ seems very low for a sgB[e] star (see text, Section \ref{section:SED}).
}
	\label{figure:spectrum-fit-best}
	\end{figure}

We will now derive the mass of the dust which is hosted in the viscous disk surrounding the sgB[e] star. By extrapolating our best-fit spectrum (see Figure \ref{figure:spectrum-fit-best}), we should have $F_\nu = 0.155$\,Jy at $60 \microns$. At this wavelength, taking an initial MRN distribution with thin ice mantles \citep[Mathis Rumpl Nordsieck;][]{mathis:1977}, the ISM dust opacity is $\kappa_\nu (60 \microns) \sim 87 \textrm{ cm}^2/\textrm{g}$ \citep{draine:2003}. We compute the black body function for $T_{\rm in} = 767$\,K, corresponding to the dust inner temperature given by the best fit (see Table \ref{table:fit}): $B(\nu,\Tdust) = 1.57 \times 10^{15}$\,Jy. Then, the total dust mass is given by $\Mdust = \frac{F_\nu \Dstar^2}{\kappa_\nu B(\nu,\Tdust)}$ where $F_\nu = 2\int_{r_{\rm in}}^{r_{\rm out}} rB(\nu,T\left(r\right))dr$, with $r_{\rm in}$ and $r_{\rm out}$ corresponding to the inner and outer radii given by the best fit, respectively equal to $0.74$ and $3.47 \frac{\rm a.u.}{\rm kpc}\sqrt(cos(i))$ (see Table \ref{table:fit}). We therefore derive a dust mass in the disk of $\Mdust \sim 5.43 \times 10^{-9} {{\Dstar}_{\rm 1\,kpc}}^2 \Msol$, where ${\Dstar}_{\rm 1\,kpc}$ is the star distance expressed in kpc. For a distance of $1.6$\,kpc \citep{rahoui:2008}, we find $\Mdust \sim 1.39 \times 10^{-8} \Msol = 2.76 \times 10^{22}$ \,kg.

This dust mass is lower than derived by \cite{moon:2007} ($3.1 \times 10^{-8} \Msol$ and $4.6 \times 10^{-10} \Msol$ for the warm and hot components respectively). This is consistent with the fact that for a same obscuration, a shell geometry will require more material than a disk geometry. This also shows that, unlike in \cite{moon:2007}, our fits do not require any hot component.


There are a few possibilities to explain the low mass of dust composing this stellar disk, compared to e.g. standard giant stars surrounded by disk, where the dust mass in the disk is more of the order of $10^{29}$\,g \citep[see e.g.][]{jura:2001,jura:2003}: {\it i.} the disk might be smaller in sgB[e] stars due to evolution effect, {\it ii.} special conditions such as X-ray ionisation might destroy the dust in the disk, and/or {\it iii.} the compact object, accreting from the stellar wind, might have an influence on the mass of dust comprised in the disk (since the compact object likely orbits inside the disk, as suggested by the high absorption and the extension of the dust).

The low mass of dust could also be explained by tidal torques created by the star, disk and compact object, and causing the disk to be  potentially truncated \citep{okazaki:2001}, likely at the $r_{\rm in}$ distance. The compact object might orbit within or at the edge of the disk, with a small inclination, quasi-coplanar with the disk, causing the extreme absorption seen in X-rays, probably local around the compact object (see Section \ref{section:introduction}). In this case the wind accretion would be fully consistent with the disk geometry derived by SED fitting. However it is difficult to conclude on this disk truncation, since the orbital separation is unknown for this system.

\begin{figure}[!ht]
	\begin{center}
	\includegraphics[angle=0,width=17cm]{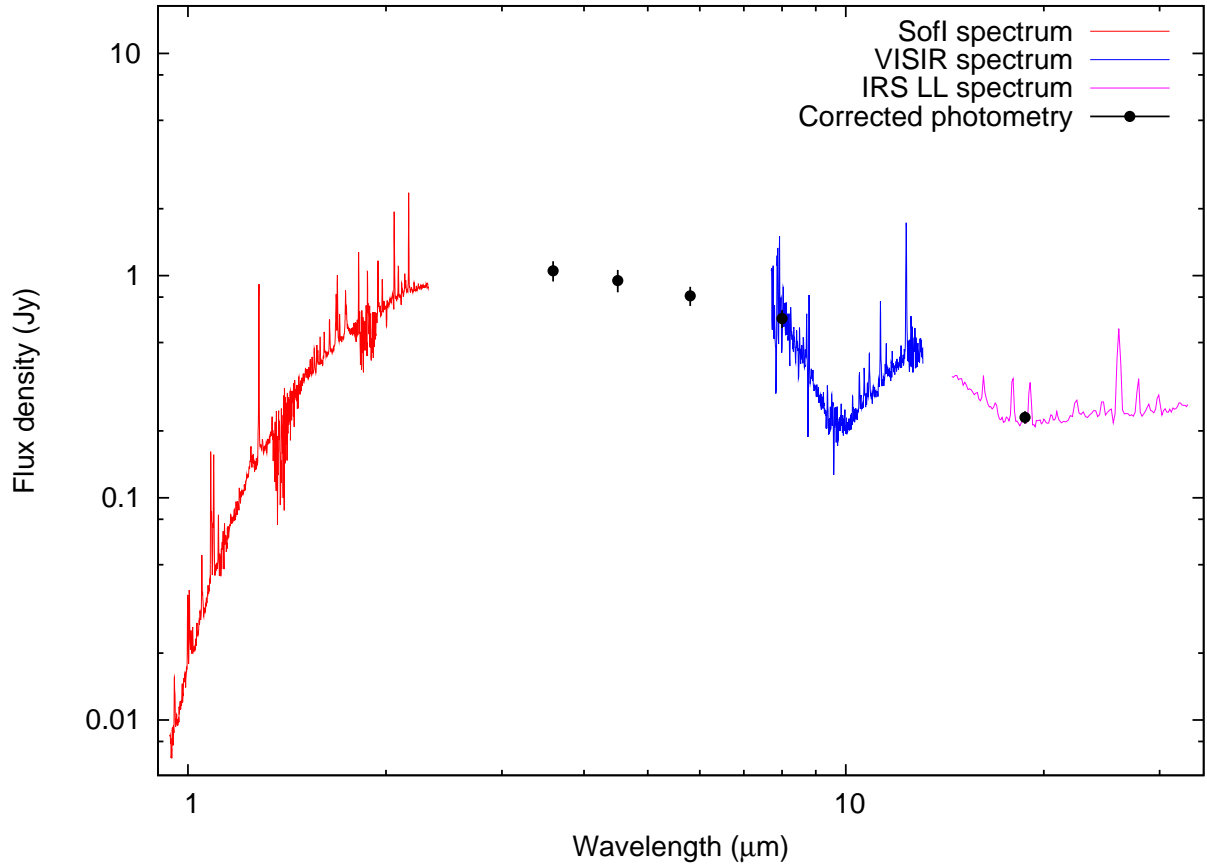}
	\end{center}
	\caption{
Broadband NIR to MIR ESO NTT/SofI, VISIR and \textit{Spitzer} {\it reddened} spectrum of $\igrjstu$, from 0.9 to $35 \microns$. The ``corrected photometry'' corresponds to VISIR + \textit{Spitzer} photometric data, taking into account the transmission factor.
}
	\label{figure:spectrumALL}
	\end{figure}

All the line ratios reported in Section \ref{section:spectroscopy} indicate an effective temperature T$_{\rm eff} \sim 35000-40000$\,K. This effective temperature is higher than the temperature of a hot B1 supergiant star, of luminosity between $10^5$ and $10^6 \Lsol$. For instance, \cite{filliatre:2004} derived a temperature of T~$= 20250$~K by fitting the optical-NIR SED. However we point out that while the NIR spectrum mainly emanates from the stellar photosphere, the MIR spectrum comes from the heated rim and the dusty viscous disk, as is shown by both the broadband and best-fit spectra (see Figures \ref{figure:spectrumALL} and \ref{figure:spectrum-fit-best}).
Therefore the location where the MIR lines are produced does not correspond to the stellar photosphere, but instead to dense and hot regions protected from the stellar radiation, and probably irradiated by X-ray emission coming from the compact object accreting from the stellar wind. In addition, this X-ray irradiation could explain the variability of a few lines, while most lines exhibit a stable flux.

We used in the computations above a distance of $1.6$\,kpc as derived by \cite{rahoui:2008}, however this is the value derived with a star dominating at NIR wavelengths, and it is possible to have a star located further away. The best fit gives $\Rstar / \Dstar = 11.71 \Rsol$\,/kpc. A distance of $1.5-2$\,kpc gives $\Rstar = 17.6-23.4 \Rsol$, corresponding to a typical radius of supergiant star. However, considering the more physical fit with $T_{\rm rim} = 5500$\,K, the best-fit parameter $\Rstar / \Dstar = 5.80 \Rsol$\,/kpc requires a distance of $3-4 \kpc$ to get a stellar radius corresponding to a supergiant star (see Table \ref{table:fit}).

In the broadband spectrum we see two additional absorption components, at around $10$ and $20 \microns$ (see Figure \ref{figure:spectrumALL}), likely corresponding to silicate absorption. We point out that these components might be due either to an absorption law which is not accurate enough in this MIR domain, or even to auto-absorption of a disk-like system that would be seen nearly edge-on, which in this case would be suggesting a high inclination system.


These observations allowed us to relate the presence of the hot dust component to the nature of the early-type star and to stellar wind photoionization. Indeed, sgB[e] stars usually exhibit hot (T$\sim$1000K) circumstellar dust, and show forbidden line emission in the MIR \citep[see e.g.][]{lamers:1998}. While the evolutionary status is not well known yet, they will likely evolve into extreme and still poorly known Luminous Blue Variables (LBV) or Wolf-Rayet (WR) stars. MIR spectroscopic characteristics of $\igrjstu$ appear very similar to those from LBVs, some of them exhibiting also thermal MIR dust emission \citep{lamers:1996a}. If the evolution from sgB[e] to LBV is confirmed, these highly obscured {\it INTEGRAL} sources might therefore be the progenitors of LBV stars, or WR stars in case of high mass loss, or even dusty wind hypergiant stars. If this is the case, then obscured {\it INTEGRAL} sources might represent a previously unknown evolutionary phase of X-ray sources hosting early-type companion stars \citep[see discussion in][]{moon:2007}. 

The question then arises: does the supergiant star influence its close environment, or does the interstellar medium influence the star, by a feedback effect? As dust is a strong tracer of star formation, the next step is to observe these sources with {\it Herschel}, since its greatest strength is the possibility to study the history of star formation in our Galaxy. This makes observations of such obscured sources with {\it Herschel} an enormously powerful tool, giving a new insight for studying the process of formation and evolution of very massive stars in molecular clouds. These observations should be valuable in the study of the later phases of stellar evolution, particularly circumstellar shells, mass-loss in general, and stellar winds, and they should also allow us to relate the properties of these objects to their birth environment. Higher wavelength observations should also allow us to better constrain the broadband SED and to confirm $T_{\rm rim}$ and $T_{\rm in}$ we derived from our observations. They would likely close the debate on the presence of colder ($<200$~K) circumstellar dust component around this source, with an extension at larger radii from the compact object of more than $30 \Rstar$, suggested by \cite{moon:2007} but not confirmed by our observations. Finally, these observations should allow us to investigate where this material comes from, and the study of such an unusual circumstellar environment should in the future tell us whether it is due to stellar evolution or to the binarity of the system itself. 

\section{Conclusion}

We have performed new VISIR photometric and spectroscopic MIR observations of the highly absorbed HMXB $\igrjstu$ hosting a supergiant B[e] star, and we have fit its NIR-MIR SED including ESO NTT/SofI, VISIR and \textit{Spitzer} data.
The main results are:

\begin{itemize}

\item the photometry did show two periods of significant variability: first an increase in PAH1 between 2005 and 2006, then a decrease in PAH2 between 2006 and 2007. Apart from this the source remained at a constant flux, within uncertainties.

\item the spectra exhibit many emission lines from various elements, some coming from forbidden atomic and ionic fine-structure lines of low ionization potential. The line diagnostics all point towards a source of temperature between 35000 and 40000\,K, likely emanating from a heated zone surrounding the stellar photosphere.

\item the multi-wavelength optical to MIR SofI-VISIR-\textit{Spitzer} SED fitting allowed us {\it i.} to exclude a spherical geometry for the dust component, and {\it ii.} by using a model adapted for sgB[e] stars, to show the presence of two-temperature components, with an irradiated rim of temperature $T_{\rm rim} = 3786$\,K, surrounded by a hot dusty viscous disk component at $T_{\rm in} = 767$~K, corresponding to the best fit parameters (see Table~\ref{table:fit}). However, the presence of \ion{O}{1} lines rather points towards irradiated rim temperatures of $\sim 5500$\,K, for which we achieved similar goodness of fits.

\end{itemize}

These results allowed us to strengthen the presence of absorbing material (dust and cold gas) enshrouding the whole binary system of $\igrjstu$, and better characterise its properties. They are therefore of prime importance to constrain the nature of this system in particular. In addition, they will also allow us to better understand the whole population of new high energy binary systems born with two very massive components, which has been recently revealed by {\it INTEGRAL}, these sources being the most extremely absorbed ones among high energy sources.
These features are most likely associated with the mass losses of OB supergiants and may play an important role in their evolutions to the final supernova explosion and formation of compact objects.
These systems are probably the primary progenitors of neutron star/black hole binary mergers, with the possibility that they are related with short/hard gamma-ray bursts, and that they might be good candidates of gravitational wave emitters. It therefore appears that, because these obscured X-ray binaries represent a different evolutionary state of X-ray binaries, their study is of prime importance, and will provide hints on the formation and evolution of high energy binary systems, and a better understanding of the evolution of such rare and short-living HMXBs.

\section{Acknowledgments}

SC would like to thank the ESO staff who were very helpful during visitor observations, Peter A. Curran for a careful rereading of the paper and useful comments, and finally the anonymous referee for constructive comments on this paper.
IRAF is distributed by the National Optical Astronomy Observatory, which is operated by the Association of Universities for Research in Astronomy (AURA) under cooperative agreement with the National Science Foundation.
This research has made use of NASA's Astrophysics Data System Bibliographic Services.  
This work was supported by the Centre National d'Etudes Spatiales (CNES). It is based on observations obtained with MINE: the Multi-wavelength {\it INTEGRAL} NEtwork. 


{\it Facilities:} \facility{ESO}.





\clearpage

\clearpage

\clearpage

\clearpage

\clearpage

\clearpage

\clearpage

\clearpage

%

\clearpage
\begin{deluxetable}{cccccc}
\tabletypesize{\scriptsize}
\tablecaption{VISIR Photometry of $\igrjstu^a$.
\label{table:photometry}}
\tablewidth{0pt}
\tablehead{
\colhead{Filters} & \colhead{$\lambda (\mu m)$} & \colhead{$\Delta \lambda (\mu m)$} & \colhead{Flux 2005 (mJy)} & \colhead{Flux 2006 (mJy)} & \colhead{Flux 2007 (mJy)}
}
\startdata
%
%
PAH1    & 8.59  & 0.42 & $409.2\pm2.4$  & $426.2\pm3.0$  & $420.98\pm3.07$  \\
ArIII   & 8.99  & 0.14 &                &                & $281.25\pm4.92$  \\
SIV\_1  & 9.82  & 0.18 &                &                & $210.93\pm14.79$ \\
SIV     & 10.49 & 0.16 &                &                & $249.63\pm4.12$  \\
SIV\_2  & 10.77 & 0.19 &                &                & $254.59\pm4.22$  \\
PAH2    & 11.25 & 0.59 & $322.4\pm3.3$  & $317.4\pm3.4$  & $290.96\pm3.01$  \\	
SiC     & 11.85 & 2.34 &                &                & $361.33\pm3.16$  \\
PAH2\_2 & 11.88 & 0.37 &                &                & $321.02\pm3.39$  \\
NeII\_1 & 12.27 & 0.18 &                &                & $372.69\pm5.75$  \\
NeII    & 12.81 & 0.21 &                &                & $376.74\pm8.31$  \\
NeII\_2 & 13.04 & 0.22 &                &                & $365.98\pm7.38$  \\
Q1      & 17.65 & 0.83 &                &                & $232.24\pm14.40$ \\
Q2      & 18.72 & 0.88 & $172.1\pm14.9$ & $180.7\pm15.3$ & $208.13\pm11.27$ \\
\enddata
\tablenotetext{a}{Photometry taken on 2005 June 21, 2006 June 30 \citep{rahoui:2008} and 2007 July 12 (this paper).}
\end{deluxetable}


\clearpage
\begin{deluxetable}{ccccccc}
\tabletypesize{\scriptsize}
\tablecaption{Detected lines in the $\igrjstu$ VISIR MIR spectrum \label{table:spectroscopy}}
\tablewidth{0pt}
\tablehead{
\colhead{Identification} & \colhead{$\lambda$ ($\mu$m)} & \colhead{$\lambda_{\rm fit}$ ($\mu$m)} & \colhead{Flux ($10^{-21}$W/cm$^2$)}  & \colhead{SNR$_{\rm fit}$} & \colhead{EW ($\mu$m)} & \colhead{FWHM ($10^{-2} \mu$m)}
}
\startdata
%
{H}{I}~(6-5)           & 7.4599 & 7.4638$\pm$2.4361$\times 10^{-4}$ & 824.08$\pm$39.37   & 30.17           &-0.51163& 1.5894$\pm$2.4401$\times 10^{-2}$ \\
$[${Ne}{VI}$]$         & 7.6524 & 7.6630$\pm$1.0785$\times 10^{-3}$   & 148.91$\pm$32.60   & 6.86            &-0.06377& 1.5871$\pm$1.1351$\times 10^{-2}$ \\
PAH	           & 7.7000 & 7.7080$\pm$5.6055$\times 10^{-4}$ & 27.67$\pm$2.74 & 15.64           &-0.00924& 1.8649$\pm$6.3628$\times 10^{-2}$ \\
{H}{I}~(16-8)          & 7.7804 & 7.7511$\pm$9.0737$\times 10^{-4}$   & 44.00$\pm$6.00   & 9.20            &-0.01771& 2.0798$\pm$9.0719$\times 10^{-2}$ \\
$[${Ar}{V}$]$?         & 7.9016 & 7.8706$\pm$1.8145$\times 10^{-3}$ & 74.64$\pm$17.16 &  4.95           &-0.03614& 2.5016$\pm$1.8166$\times 10^{-2}$ \\ 
$[${Ar}{V}$]$?         & 7.9016 & 7.9274$\pm$3.8353$\times 10^{-4}$ & 43.67$\pm$3.72 & 17.95           &-0.01898& 1.3931$\pm$3.8038$\times 10^{-2}$ \\
{He}{II}~(21-14)/H$_2$ & 8.0385 & 8.0309$\pm$9.4853$\times 10^{-4}$ & 16.26$\pm$3.30 &  7.53           &-0.00764& 1.4821$\pm$9.7625$\times 10^{-2}$ \\
{He}{II}~(24-15)p-     & 8.4128 & 8.3882$\pm$1.0484$\times 10^{-3}$ & 5.31$\pm$0.90 &  7.73           &+0.00330& 1.9228$\pm$1.0491$\times 10^{-2}$ \\
{He}{II}~(24-15)p+     & 8.4128 & 8.4187$\pm$5.7743$\times 10^{-4}$ & 3.63$\pm$1.05 &  9.38           &-0.00248& 0.95614$\pm$9.3446$\times 10^{-2}$ \\
{H}{I}~(25-9)p-        & 8.4849 & 8.4772$\pm$1.1904$\times 10^{-3}$ & 7.03$\pm$1.42 &  6.66           &+0.00499& 1.8254$\pm$1.1948$\times 10^{-2}$ \\	
{H}{I}~(25-9)p+        & 8.4849 & 8.5066$\pm$7.8518$\times 10^{-4}$ & 5.50$\pm$2.52 &  6.48           &-0.00466& 0.95303$\pm$1.4639$\times 10^{-2}$ \\
PAH                & 8.6000 & 8.6216$\pm$1.1826$\times 10^{-3}$ & 7.70$\pm$1.34 &  7.10           &-0.00534& 2.1248$\pm$1.1826$\times 10^{-2}$ \\		
{H}{I}~(14-8)          & 8.6645 & 8.6844$\pm$1.8506$\times 10^{-3}$   & 9.46$\pm$1.82   & 6.95            &-0.00715& 3.1814$\pm$2.2038$\times 10^{-2}$ \\
{H}{I}~(23-9)          & 8.7206 & 8.7196$\pm$6.5877$\times 10^{-4}$	& 22.99$\pm$1.97   & 13.84           &-0.02097& 2.4195$\pm$6.7063$\times 10^{-2}$ \\
{H}{I}~(10-7)p-        & 8.7601 & 8.7610$\pm$7.1931$\times 10^{-4}$ & 10.86$\pm$1.29 & 11.21           &+0.01003& 1.8906$\pm$7.2333$\times 10^{-2}$ \\			
{H}{I}~(10-7)p+        & 8.7601 & 8.7916$\pm$2.5839$\times 10^{-4}$ & 18.98$\pm$1.12 & 25.52           &-0.01857& 1.3378$\pm$2.4715$\times 10^{-2}$ \\
{H}{I}~(22-9)          & 8.8697 & 8.8726$\pm$3.2955$\times 10^{-3}$ & 3.25$\pm$1.04 &  3.28	       &-0.00320& 3.3951$\pm$3.5891$\times 10^{-2}$ \\
{He}{II}~(23-15)       & 8.9209 & 8.9139$\pm$5.8644$\times 10^{-4}$ & 1.88$\pm$0.26 & 11.25           &-0.00189& 1.2705$\pm$5.6207$\times 10^{-2}$ \\
$[${Ar}{III}$]$        & 8.9914 & 8.9838$\pm$1.8778$\times 10^{-3}$   & 2.02$\pm$0.35   & 7.43            &-0.00191& 3.4057$\pm$2.1413$\times 10^{-2}$ \\
$[${Na}{IV}$]$         & 9.04100& 9.0354$\pm$6.6001$\times 10^{-4}$   & 1.65$\pm$0.23   & 10.62           &-0.00206& 1.4531$\pm$6.6341$\times 10^{-2}$ \\
{H}{I}~(20-9)          & 9.2605 & 9.2450$\pm$7.8913$\times 10^{-4}$ & 2.32$\pm$0.36 &  9.44           &-0.00339& 1.6305$\pm$8.0929$\times 10^{-2}$ \\
{H}{I}~(13-8)          & 9.3920 & 9.3791$\pm$1.0614$\times 10^{-3}$ & 3.85$\pm$0.80 &  7.09           &-0.00723& 1.5337$\pm$1.0270$\times 10^{-2}$ \\
{H}{I}~(19-9)          & 9.5217 & 9.5197$\pm$2.0793$\times 10^{-3}$ & 4.91$\pm$1.47 &  4.36           &-0.00864& 2.2634$\pm$2.2532$\times 10^{-2}$ \\
{He}{II}~(13-11)       & 9.7068 & 9.7011$\pm$1.2559$\times 10^{-3}$ & 1.68$\pm$0.50 &  5.25           &-0.00315& 1.2286$\pm$1.1709$\times 10^{-2}$ \\
{H}{I}~(18-9)          & 9.8470 & 9.8385$\pm$1.8978$\times 10^{-3}$ & 2.46$\pm$0.64 &  4.68           &-0.00556& 2.2627$\pm$1.8980$\times 10^{-2}$ \\
{H}{I}~(17-9)          &10.2613 &10.2434$\pm$2.3518$\times 10^{-3}$ & 3.95$\pm$1.07 &  4.29           &-0.00804& 2.8380$\pm$2.5516$\times 10^{-2}$ \\
{H}{I}~(12-8)/{He}{II}~(24-16) &10.4992 &10.4871$\pm$5.5642$\times 10^{-4}$ & 7.50$\pm$0.44 & 20.79           &-0.01486& 2.9583$\pm$5.5727$\times 10^{-2}$ \\
$[${S}{IV}$]$/$[${Co}{II}$]$? & 10.5210 &10.5424$\pm$4.9146$\times 10^{-4}$ & 1.11$\pm$0.52 & 14.62           &-0.00224& 1.0418$\pm$1.5577$\times 10^{-2}$ \\ 
$[${Ni}{II}$]$     &10.6822 &10.6662$\pm$1.1298$\times 10^{-3}$ & 6.45$\pm$0.83 &  9.89	       &-0.01386& 2.7287$\pm$1.1256$\times 10^{-2}$ \\
{H}{I}~(16-9)      &10.8036 &10.7851$\pm$1.8913$\times 10^{-3}$ & 2.39$\pm$0.61 &  5.44           &-0.00462& 2.2775$\pm$1.8887$\times 10^{-2}$ \\
{H}{I}~(25-10)     &10.8514 &10.8588$\pm$7.4041$\times 10^{-4}$ & 10.21$\pm$0.74 & 16.24           &-0.01979& 3.1739$\pm$7.4205$\times 10^{-2}$ \\
{H}{I}~(23-10)     &11.2429 &11.2206$\pm$1.3047$\times 10^{-3}$ & 1.66$\pm$0.24 &  5.39           &-0.00237& 3.0392$\pm$1.4575$\times 10^{-2}$ \\
{H}{I}~(9-7)/{He}{II}~(23-16) &11.3026 &11.2895$\pm$7.4253$\times 10^{-4}$ & 22.33$\pm$1.85 & 15.21           &-0.04125& 2.7936$\pm$7.4417$\times 10^{-2}$ \\
{H}{I}~(22-10)/$$[$$CaV$$]$$? & 11.48200 &11.4801$\pm$1.7865$\times 10^{-3}$ & 1.59$\pm$0.49 &  5.40           &-0.00257& 1.8401$\pm$1.9194$\times 10^{-2}$ \\ 
{H}{I}~(15-9)      &11.5395 &11.5197$\pm$4.6056$\times 10^{-4}$ & 5.37$\pm$0.35 & 21.94           &-0.00799& 2.2187$\pm$4.7369$\times 10^{-2}$ \\
{He}{II}~(20-15)   &11.7179 &11.7268$\pm$1.5285$\times 10^{-3}$ & 1.85$\pm$0.51 &  5.98           &-0.00304& 1.7858$\pm$1.6108$\times 10^{-2}$ \\
{H}{I}~(21-10)     &11.7914 &11.7675$\pm$1.0107$\times 10^{-3}$ & 3.93$\pm$0.59 & 10.03           &-0.00605& 2.1741$\pm$1.0708$\times 10^{-2}$ \\
{H}{I}~(20-10)     &12.1568 &12.1338$\pm$4.4487$\times 10^{-4}$ & 3.50$\pm$0.19 & 24.69           &-0.00546& 2.5959$\pm$4.6772$\times 10^{-2}$ \\
{H}{I}(7-6)/{H}{I}(11-8)/{He}{II}(14-12) &12.3669 &12.3541$\pm$5.9849$\times 10^{-4}$ & 70.95$\pm$3.76 & 26.25           &-0.12208& 3.5277$\pm$6.0072$\times 10^{-2}$ \\
{H}{I}~(14-9)      &12.5837 &12.5676$\pm$1.3871$\times 10^{-3}$ & 14.57$\pm$1.72 &  9.39           &-0.02618& 3.6829$\pm$1.4051$\times 10^{-2}$ \\
{H}{I}~(19-10)     &12.6110 &12.6248$\pm$8.5191$\times 10^{-4}$ & 6.16$\pm$0.85 & 10.99           &-0.01084& 1.9122$\pm$8.4719$\times 10^{-2}$ \\
PAH	       &12.7000 &12.6986$\pm$1.2823$\times 10^{-3}$ & 7.76$\pm$0.85 & 11.18           &-0.01320& 3.8021$\pm$1.4296$\times 10^{-2}$ \\
$[${Ni}{II}$]$     &12.7288 &12.7362$\pm$4.4682$\times 10^{-4}$ & 5.73$\pm$0.41 & 22.00           &-0.00813& 2.0217$\pm$4.7863$\times 10^{-2}$ \\
$[${Ne}{II}$]$     &12.81355&12.8281$\pm$8.1494$\times 10^{-4}$ & 2.16$\pm$0.32 & 10.87           &-0.00398& 1.7358$\pm$8.1869$\times 10^{-2}$ \\
{H}{I}~(18-10)     &13.1880 &13.1901$\pm$1.6532$\times 10^{-3}$ & 3.77$\pm$1.92 &  5.38           &-0.00332& 1.1475$\pm$1.9459$\times 10^{-2}$ \\
\enddata
\end{deluxetable}

\clearpage
\begin{deluxetable}{cccccccccc}
\tabletypesize{\scriptsize}
\tablecaption{Fitted parameters of $\igrjstu$ SofI-VISIR-\textit{Spitzer} SED$^a$.
\label{table:fit}}
\tablewidth{0pt}
\tablehead{
\colhead{Model} & \colhead{T$_*$~(K)} & \colhead{$\frac{{\rm R}_*}{{\rm D}_*}$} & \colhead{T$_{\rm rim}$}  & \colhead{h$_{\rm rim}$} & \colhead{T$_{\rm in}$~(K)} & \colhead{$r_{\rm in}$} & \colhead{$r_{\rm out}$}  & \colhead{$q$} & \colhead{$\chi^2$/dof} \\
 & fixed & $\frac{\Rsol}{\rm kpc}$ & K & $\frac{\rm a.u.}{\rm kpc}\sqrt(cos(i))$ & K & $\frac{\rm a.u.}{\rm kpc}\sqrt(cos(i))$ & $\frac{\rm a.u.}{\rm kpc}\sqrt(cos(i))$ & fixed &
}
\startdata
Flaring disk  & 20000  & 11.72$\pm$0.55 & 3791$\pm$253 & 0.33$\pm$0.22 & 1716$\pm$535 & 0.08$\pm$0.05 & 3.28$\pm$0.15 & 0.5  & 1.20 (1259) \\ \hline
Viscous disk & 20000  & 11.71$\pm$0.49 & 3786$\pm$219 & 0.037$\pm$0.002 & 767$\pm$32 & 0.74$\pm$0.06 & 3.47$\pm$0.10 & 0.75  & 1.19 (1259) \\
Viscous disk & 20000  & 11.20$\pm$0.06 & 4000 (fixed)     & 0.037$\pm$0.002 & 790$\pm$19 & 0.70$\pm$0.03 & 3.49$\pm$0.07 & 0.75  & 1.20 (1260) \\
Viscous disk & 20000  & 9.86$\pm$0.08  & 4500 (fixed)     & 0.036$\pm$0.001 & 834$\pm$20 & 0.64$\pm$0.03 & 3.50$\pm$0.08 & 0.75  & 1.19 (1260) \\
Viscous disk & 20000  & 8.16$\pm$0.12  & 5000 (fixed)     & 0.035$\pm$0.001 & 868$\pm$21 & 0.60$\pm$0.03 & 3.51$\pm$0.07 & 0.75  & 1.20 (1260) \\
Viscous disk & 20000  & 5.80$\pm$0.20  & 5500 (fixed)     & 0.035$\pm$0.001 & 895$\pm$21 & 0.57$\pm$0.02 & 3.53$\pm$0.07 & 0.75  & 1.21 (1260) \\
Viscous disk & 20000  & 0 (fixed)    & 5963$\pm$35  & 0.035$\pm$0.001 & 912$\pm$22 & 0.56$\pm$0.02 & 3.55$\pm$0.07 & 0.75  & 1.22 (1260) \\
Viscous disk & 20000  & 0            & 6000 (fixed)     & 0.035$\pm$0.001 & 923$\pm$20 & 0.55$\pm$0.02 & 3.55$\pm$0.07 & 0.75  & 1.22 (1260) \\ \hline
2-sphere      &        &                & Tdust$_1$ (K) & $\frac{{\rm Rdust}_1}{{\rm D}_*}$  & Tdust$_2$ (K)           & $\frac{{\rm Rdust}_2}{{\rm D}_*}$       & - & -  &  \\
component     & 20000  & 13.25$\pm$0.10 & 2668$\pm$43   & 0.32$\pm$0.01                      & 379$\pm$3               & 3.02$\pm$0.02 & - & -  & 1.29 (1260) \\ \hline
\enddata
\tablenotetext{a}{The best fit with SofI-VISIR-\textit{Spitzer} data is obtained for the viscous disk model, for $T_{\rm rim} = 3786$\,K, as shown in Figure \ref{figure:spectrum-fit-best}. However, this value being low for a sgB[e] star, we also tried various fits with frozen $T_{\rm rim}$ going from 4000 to 6000\,K. The fit, which physically better describes a sgB[e] star with $T_{\rm rim} \sim 5500$\,K, is reported in Figure \ref{figure:spectrum-fit-visc-5500}.
Since fitting with the 2-sphere component model gives a higher $\chi^2_r$ than fitting with the flaring disk or viscous disk models, the SofI-VISIR-\textit{Spitzer} data allow us to exclude the spherical repartition of the dust surrounding $\igrjstu$.
}
\end{deluxetable}

\begin{deluxetable}{cccccccccc}
\tabletypesize{\scriptsize}
\tablecaption{Fitted parameters of $\igrjstu$ SofI-\textit{Spitzer} SED$^a$.
\label{table:fit-Spitzer}}
\tablewidth{0pt}
\tablehead{
\colhead{Model} & \colhead{T$_*$~(K)} & \colhead{$\frac{{\rm R}_*}{{\rm D}_*}$} & \colhead{T$_{\rm rim}$}  & \colhead{h$_{\rm rim}$} & \colhead{T$_{\rm in}$~(K)} & \colhead{$r_{\rm in}$} & \colhead{$r_{\rm out}$}  & \colhead{$q$} & \colhead{$\chi^2_r$ (dof)} \\
 & fixed & $\frac{\Rsol}{\rm kpc}$ & K & $\frac{\rm a.u.}{\rm kpc}\sqrt(cos(i))$ & K & $\frac{\rm a.u.}{\rm kpc}\sqrt(cos(i))$ & $\frac{\rm a.u.}{\rm kpc}\sqrt(cos(i))$ & fixed &
}
\startdata
Flaring disk  & 20000  & 12.22$\pm$0.38 & 3538$\pm$180 & 0.038$\pm$0.006 & 629$\pm$48        & 0.80+/0.17    & 3.08$\pm$0.35  & 0.5   & 1.30 (994) \\ \hline
Viscous disk  & 20000  & 12.23$\pm$0.45 & 3532$\pm$177 & 0.022$\pm$0.002 & 587$\pm$31        & 1.37$\pm$0.18 & 3.29$\pm$0.23  & 0.75  & 1.30 (994) \\
Viscous disk  & 20000  & 11.17$\pm$0.06 & 4000 (fixed) & 0.023$\pm$0.002 & 640$\pm$19        & 1.14$\pm$0.08 & 3.22$\pm$0.20  & 0.75  & 1.31 (995) \\
Viscous disk  & 20000  & 9.79$\pm$0.08  & 4500 (fixed) & 0.023$\pm$0.001 & 679$\pm$18        & 1.01$\pm$0.06 & 3.20$\pm$0.18  & 0.75  & 1.32 (995) \\
Viscous disk  & 20000  & 8.05$\pm$0.12  & 5000 (fixed) & 0.023$\pm$0.01  & 707$\pm$19        & 0.93$\pm$0.05 & 3.19$\pm$0.16  & 0.75  & 1.33 (995) \\
Viscous disk  & 20000  & 5.59$\pm$0.20  & 5500 (fixed) & 0.023$\pm$0.01  & 729$\pm$19        & 0.88$\pm$0.05 & 3.19$\pm$0.15  & 0.75  & 1.34 (995) \\
Viscous disk  & 20000  & 0 (fixed)      & 5922$\pm$33  & 0.023$\pm$0.01  & 742$\pm$19        & 0.85$\pm$0.05 & 3.19$\pm$0.4   & 0.75  & 1.35 (995) \\
Viscous disk  & 20000  & 0              & 6000 (fixed) & 0.023$\pm$0.01  & 759$\pm$18        & 0.82$\pm$0.04 & 3.19$\pm$0.15  & 0.75  & 1.36 (995) \\ \hline
2-sphere      &        &                & Tdust$_1$ (K)& $\frac{{\rm Rdust}_1}{{\rm D}_*}$   & Tdust$_2$ (K)            & $\frac{{\rm Rdust}_2}{{\rm D}_*}$ & - & -  &  \\
component     & 20000  & 13.03$\pm$0.17 & 3108$\pm$84  & 0.27$\pm$0.01   & 412$\pm$3        & 2.82$\pm$0.02 & - & -  & 1.32 (995) \\ \hline
\enddata
\tablenotetext{b}{The fits with SofI-\textit{Spitzer} data have a higher $\chi^2_r$ than the fits including SofI-VISIR and \textit{Spitzer} data, and do not allow to exclude the 2-sphere component model.}
\end{deluxetable}
\end{document}